\documentclass[11pt,showpacs,preprintnumbers,superscriptaddress,amsmath,amssymb,nofootinbib]{revtex4}
\usepackage{graphicx}% Include figure files
\usepackage{dcolumn}% Align table columns on decimal point
\usepackage{bm}% bold math
\usepackage{amssymb}
\usepackage{amsmath}
\usepackage{epsfig}    
\usepackage{color}
\usepackage{slashed}
\usepackage{hhline}
\usepackage{ulem}
\def\be{\begin{equation}}
\def\ee{\end{equation}}
\newcommand{\bea}{\begin{eqnarray}}
\newcommand{\eea}{\end{eqnarray}}
\newcommand{\nn}{\nonumber}

\begin{document}

%{\begin{flushright}{APCTP Pre2023 - 0XX}\end{flushright}}

%%%%%%%%%
\title{A new type of lepton seesaw model in a modular $A_4$ symmetry}

%\preprint{KYUSHU-HET-268}

\author{Takaaki Nomura}
\email{nomura@scu.edu.cn}
\affiliation{College of Physics, Sichuan University, Chengdu 610065, China}

\author{Hiroshi Okada}
\email{hiroshi3okada@htu.edu.cn}
\affiliation{Department of Physics, Henan Normal University, Xinxiang 453007, China}

\date{\today}

\begin{abstract}
{We propose a new type of lepton seesaw model introducing a modular $A_4$ flavor symmetry in which isospin doublet vector fermions play an important role in constructing seesaw mechanisms for both the charged-lepton mass matrix and the neutrino one.
The charged-lepton mass matrix is induced via the Dirac seesaw with a two-by-two block mass matrix.
On the other hand, the neutrino mass matrix is generated via a seesaw with four-by-four block mass matrix where the right-handed neutral fermions are also added.
{Remarkably the neutral fermion mass matrix provides us a new type of seesaw formula for active neutrino mass.}
Through our chi square analysis, we find some tendencies about observables focusing on two fixed points $\tau=i,\ \omega$.
}
%$A_4$
%%%%%%%%%%%%%%%%%%%%%%%%%%
 %
 \end{abstract}
\maketitle
\newpage

\section{Introduction}
Understanding a mechanism of non-vanishing neutrino masses is one of the most important issues beyond the standard model (BSM).
Due to neutral under the electric charge, it would be natural to assume that the neutrino is Majorana fermions~\footnote{If one assumes the neutrino is Dirac fermion, one has to impose an additional symmetry to forbid the Majorana nature.}.
In general, we need to add BSM fields to generate neutrino masses and explain its smallness. 
{In addition to the explanation of neutrinos, some phenomenological issues require new physics such as an existence of dark matter, muon anomalous magnetic dipole moment~\cite{Muong-2:2023cdq}, W boson mass anomaly~\cite{CDF:2022hxs},
some anomalies in semi-leptonic decays of mesons~\cite{London:2021lfn,Iguro:2022yzr}, and so on.
Vector-like fermions would be one of the most promising candidates to resolve these phenomenological issues.
In fact vector-like fermions are applied, for example, to explain
DM~\cite{Cirelli:2005uq,Bhattacharya:2015qpa,Baek:2017ykw}, muon $g-2$~\cite{Poh:2017tfo, Kawamura:2019rth, Nomura:2022wck, Nagao:2022oin, Lee:2022nqz}, W boson mass anomaly~\cite{Nagao:2022oin, Lee:2022nqz}, semi-leptonic meson decays~\cite{Kawamura:2017ecz, Poh:2017tfo,Kawamura:2019rth, Ko:2021lpx}, and neutrino mass matrix~\cite{Okada:2015nca}.
Furthermore vector-like fermions are considered in association with flavor physics~\cite{Ishiwata:2015cga,Poh:2017tfo}, electroweak symmetry breaking vacuum~\cite{Xiao:2014kba} and the Higgs boson physics~\cite{delAguila:1989rq, Ellis:2014dza,Arhrib:2016rlj, Chen:2017hak}.
Thus, it is interesting to explore an application of vector-like fermions.}
%
%\footnote{See following papers for example that discuss DM~\cite{Cirelli:2005uq,Baek:2017ykw}, muon $g-2$,~\cite{Nomura:2022wck, Nagao:2022oin, Lee:2022nqz}, W boson mass anomaly,~\cite{Nagao:2022oin, Lee:2022nqz}, semi-leptonic decays~\cite{Ko:2021lpx}, and neutrino mass matrix~\cite{Okada:2015nca}, applying the vector-like fermions.}.
%%% If possible the references wcould be better to be restriced by su(2) 'doublet' vector-fermions. %%%%%%%%

In this paper, we propose a new type of seesaw mechanism introducing three isospin doublet vector-like fermions %in addition to 
and Majorana right-handed heavy neutral fermions.
The vector-like fermions also contribute to the lepton mass matrix  which is induced via seesaw-like mechanism(lepton seesaw)~\cite{Lee:2021gnw, Nomura:2024ctl}.
{Especially we realize a new type of seesaw formula in the generation of active neutrino mass in which cubic of heavy fermion masses appear in the denominator.}
In order to realize our new mechanism, we apply a modular $A_4$ flavor symmetry which might be the minimum symmetry in fact.
\footnote{We have checked that any other gauged symmetries cannot be applied to construct our model. 
If one adopts an global symmetry, it is possible. But Higgs has to have nonzero charge under the new symmetry. It might cause a danger Goldstone boson that would affect Higgs interactions and it would be ruled out by collider physics.}
The modular $A_4$ flavor symmetry is frequently applied to quark and lepton sectors in order to explain/predict their masses and mixings and CP phases,
and it is a prominent symmetry to control some desired or undesired terms due to a unique nature of modular weight~\cite{Feruglio:2017spp, Criado:2018thu, Kobayashi:2018scp, Okada:2018yrn, Nomura:2019jxj, Okada:2019uoy, deAnda:2018ecu, Novichkov:2018yse, Nomura:2019yft, Okada:2019mjf, Ding:2019zxk, Nomura:2019lnr,Kobayashi:2019xvz,Asaka:2019vev,Zhang:2019ngf, Gui-JunDing:2019wap,Kobayashi:2019gtp,Nomura:2019xsb, Wang:2019xbo,Okada:2020dmb,Okada:2020rjb, Behera:2020lpd, Behera:2020sfe, Nomura:2020opk, Nomura:2020cog, Asaka:2020tmo, Okada:2020ukr, Nagao:2020snm, Okada:2020brs,Kang:2022psa, Ding:2024fsf, Ding:2023htn, Nomura:2023usj, Kobayashi:2023qzt, Petcov:2024vph, Kobayashi:2023zzc, Nomura:2024ghc, Nomura:2024ctl, Nomura:2024vus, Baur:2024lcc, Ding:2024euc}.~\footnote{Recently, non-holomorphic modular symmetries are arisen by Qu and Ding group~\cite{Qu:2024rns} and several papers are appeared after the paper~\cite{Nomura:2024vzw, Nomura:2024atp, Ding:2024inn, Li:2024svh, Nomura:2024nwh, Okada:2025jjo, Kobayashi:2025hnc, Loualidi:2025tgw}.}

This paper is organized as follows.
In Sec. \ref{sec:II}, 
we show how to construct our model applying the modular $A_4$ symmetry with new particles and discuss mechanisms of the charged-lepton sector, and neutral fermion sector including the active neutrinos. 
In Sec. \ref{sec:III}, we carry out the chi-square analyses at nearby two fixed points $\tau=i,\ \omega$ 
and demonstrate tendencies of favorite allowed ranges.
Finally, we summarize and conclude in Sec. \ref{sec:IV}.

\section{Model setup}
\label{sec:II}

\begin{table}[t!]
\begin{tabular}{|c||c|c|c|c|c||c|c|c|}\hline\hline  
& ~$\overline{\ell}$~ & ~$L$~& ~$ \overline{L'}$~& ~$ {L'}$~& ~$ \overline{N}$~ & ~$ H_u$ ~&~ {$H_d$}~~&~ {$\varphi$}~  \\\hline\hline 
%%%
$SU(2)_L$   & $\bm{1}$ & $\bm{2}$   & $\bm{2}$ & $\bm{2}$ & $\bm{1}$  & $\bm{2}$ & $\bm{2}$    & $\bm{1}$  \\\hline 
$U(1)_Y$     & $1$ & $-\frac12$  & $\frac12$ & $-\frac12$  & $0$ & $\frac12$ & $-\frac12$  & $0$    \\\hline
$A_4$   & $ \{ \bar{\bm{1}} \}$ & $\bm{3}$  & $\{ \bar{\bm{1}} \}$ & $\{ \bm{1} \} $  & $\bm{3}$ & $\bm{1}$ & $\bm{1}$ & $\bm{1}$         \\\hline 
$-k_I$  & $-2$  & $+3$  & $+2$ & $-2$& $-2$ & $0$ & $0$  & $-7$     \\\hline
\end{tabular}
\caption{Charge 
assignments of the fermions and bosons
under $SU(2)_L\otimes U(1)_Y \otimes A_4$ where $-k_I$ is the number of modular weight. Here, $\{ \bm{1} \} =\{1, 1', 1''\}$
and  $\{ \bar{\bm{1}} \} =\{1, 1'', 1'\}$ indicates assignment of $A_4 $singlets.} 
\label{tab:1}
\end{table}
In this section, we introduce our model and show mass matrices of charged/neutral leptons.
In the SM lepton sector, we respectively assign $\overline{\ell}$ and $L(\equiv[\nu,\ell]^T)$ to be $\{ \bar{\bm{1}} \} =\{1, 1'', 1'\}$ and $\bm{3}$ under the $A_4$ symmetry where each of the modular weight is distributed to $-2$ and $+3$.
In addition, we introduce three families of vector-like leptons $L'(\overline{L'})$, which is denoted by $[N',E']^T([\overline{N'},\overline{E'}]^T)$, assigned by $\{{\bm{1}} \} (\{ \bar{\bm{1}} \})$ with $-2(+2)$ modular weight. Three families of neutral Majorana fermions $\overline{N}$ are added under the $A_4$ triplet with $-2$ modular weight.
%%%
In the Higgs sector, we introduce a singlet scalar $\varphi=(v_\varphi + r_\varphi+i z_\varphi)/\sqrt2$ with trivial singlet under $A_4$ with $-7$ modular weight in addition to the
two Higgs doublets $H_u=[h^+_u,(v_u+ h_u+iz_u)/\sqrt2]^T$ and $H_d=[(v_d+ h_d+iz_d)/\sqrt2, h^-_d]^T$ in the MSSM.
Here, $H_u$ and $H_d$ are trivial $A_4$ singlet under zero modular weight.
Due to these assignments, we forbid terms $\overline{\ell} L H_d$ and $\overline{N} L H_u$, which would give a standard lepton sector with the canonical seesaw model. 
$\overline{N} \overline{L'} H_d$ and  $\overline{L'}{L'} \varphi$ are also forbidden in order to get our desired neutrino mass matrix as can be seen later.

\subsection{Charged-fermion mass matrix}
The mass matrix of charged-fermion is arisen from the following superpotential that is under the modular $A_4$ symmetry; $ \overline{\ell} \otimes H_d\otimes L',\ Y^{(2)}_3\otimes \overline{L'}\otimes L\otimes\varphi ,\  \overline{L'}\otimes L'$.
Here, $Y^{(2)}_3\equiv[y_1,y_2,y_3]^T$ is  the $A_4$ triplet modular Yukawa with 2 modular weight~\footnote{The component of modular form $y_{1,2,3}$ is written by a function of modulus $\tau$ (details are referred to ref.~\cite{Feruglio:2017spp}). The other modular forms are constructed by the products of weight 2 modular form.}.
After the spontaneous electroweak symmetry breaking, the relevant terms to induce the charged-lepton mass matrix are explicitly given by $ \overline{\ell}  m E',\ \overline{E'} m' \ell ,\  \overline{E'} M_{L'} E$
where each form of $m$, $m'$, and $M_{L'}$ is found as
\begin{align}
m = \frac{v_d}{\sqrt2}
 \left(\begin{array}{ccc} 
a_e & 0 & 0 \\
0 & b_e& 0\\
0& 0 & c_e  \end{array} \right),\
m' = \frac{{v_\varphi}}{\sqrt2}
 \left(\begin{array}{ccc} 
f_1 & 0 & 0 \\
0 & f_2 & 0\\
0& 0 & f_3  \end{array} \right)
 \left(\begin{array}{ccc} 
y_1 & y_3 & y_2 \\
y_2 & y_1 & y_3\\
y_3& y_2 & y_1  \end{array} \right),\
M_{L'}=  
 \left(\begin{array}{ccc} 
M_1 & 0 & 0 \\
0 & M_2 & 0\\
0& 0 & M_3  \end{array} \right),
\label{eq:cgd-lep2}
\end{align}
where $\{a_e, b_e, c_e \}$, $\{f_1,f_2,f_3\}$, and $\{M_1,M_2,M_3\}$ can be real parameters without loss of generality.
%%%
For our convenience of the neutrino analysis, we redefine $m' \equiv \frac{{v_\varphi} f_1}{\sqrt2}\tilde{m}'$ and $M_{L'} \equiv M_1  \tilde{M}_{L'}$.
%%%
Then, the mass matrix of charged-lepton sector ${\cal M}_e$ is given in terms of two by two block matrix as follows:
\begin{align}
%m' = \frac{v'}{\sqrt2}{\cal M}_e=
 \left(\begin{array}{c} 
\overline{\ell}  \\
\overline{E'}
  \end{array} \right)^T
 \left(\begin{array}{cc} 
0 & m \\
m' &M_{L'}
  \end{array} \right)
 \left(\begin{array}{c} 
{\ell}  \\
{E'}
  \end{array} \right) .
\label{eq:cgd-mtrx}
\end{align}
Assuming $m,m' \ll M_{L'}$,
we compute the charged-lepton mass matrix as a seesaw-like as follows.
First of all, let us suppose ${\cal M}_e$ can be diagonalized by
\begin{align}
%m' = \frac{v'}{\sqrt2}
\left(\begin{array}{cc} 
m_\ell & 0 \\
0 &M_{\ell}
  \end{array} \right)= 
  O^\dag_R{\cal M}_e O_L.
% \label{eq:cgd-mtrx}
\end{align}
Then, we analytically write the mixing matrix and mass matrices as follows:
\begin{align}
& O_{L}\simeq
\left(\begin{array}{cc} 
{\bf 1}_{3\times3} & M^{-1}_{L'} m'\\
-M^{-1}_{L'} m' & {\bf 1}_{3\times3} 
  \end{array} \right)
   %\theta_{L}\approx -M^{-1}_{L'} m'
 ,\nn\\
 %%%
& m_\ell \approx m M_{L'}^{-1} m',\quad M_\ell  \approx M_{L'},
 \label{eq:cgd-mtrx}
\end{align}
where $m,m' \ll M_{L'}$ is applied.
%Since $M_{L'}$ is diagonal, we do not need to diagonalize it more.
Furthermore, $m_\ell $ is diagonalized by $D_e=V_{e_R}^\dag m_\ell  V_{e_L}$;
we obtain $|D_e|^2=V^\dag_{e_L} m^\dag_\ell  m_\ell  V_{e_L}$ where $D_e=$
diag.$(m_e,m_\mu,m_\tau)$.
%%%
Using Eqs.~(\ref{eq:cgd-lep2}) and (\ref{eq:cgd-mtrx}), 
$m_\ell $ is written down as 
\begin{align}
 \frac{v_d {v_\varphi} }{2}
 \left(\begin{array}{ccc} \frac{a_e f_1}{M_1} & 0 & 0 \\
0 & \frac{b_e f_2}{M_2} & 0\\
0& 0 & \frac{c_e f_3}{M_3}  \end{array} \right)
 \left(\begin{array}{ccc} 
y_1 & y_3 & y_2 \\
y_2 & y_1 & y_3\\
y_3& y_2 & y_1  \end{array} \right).
\label{eq:me}
\end{align}
Once concrete value of $\tau$ is given in the fundamental region, $y_{1,2,3}$ is uniquely fixed. Furthermore, parameters $f_{1,2,3}$ and $M_{1,2,3}$ contribute to explanation of the active neutrino mass matrix. 
Thus, $\{a_e, b_e, c_e\}$ are free input  parameters to fit the experimental values $\{m_e,m_\mu,m_\tau\}$ by the following relations:
\begin{align}
&{\rm Tr}[m^\dag_e m_e] = |m_e|^2 + |m_\mu|^2 + |m_\tau|^2,\quad
 {\rm Det}[m^\dag_e m_e] = |m_e|^2  |m_\mu|^2  |m_\tau|^2,\nn\\
&({\rm Tr}[m^\dag_e m_e])^2 -{\rm Tr}[(m^\dag_e m_e)^2] =2( |m_e|^2  |m_\mu|^2 + |m_\mu|^2  |m_\tau|^2+ |m_e|^2  |m_\tau|^2 ).\label{eq:l-cond}
\end{align}

\subsection{Neutral fermion mass matrix}
%In this subsection, 
The mass matrix of neutral fermion is arisen from the following superpotential that is under the modular $A_4$ symmetry;
$Y^{(4)}_{3}\otimes \overline{N} \otimes H_u \otimes L',\ Y^{(2)}_3\otimes \overline{L'}\otimes L\otimes\varphi ,\  \overline{L'}\otimes L',\ Y^{(4)}_{1,1',3}\otimes\overline{N}\otimes C \overline{N}^T\ (C\equiv i\gamma_2\gamma_0)$.
Here, $Y^{(4)}_3\equiv[y'_1,y'_2,y'_3]^T$ is  the $A_4$ triplet modular Yukawa with 4 modular weight.
After the spontaneous electroweak symmetry breaking, the relevant terms to induce the neutral fermion mass matrices are explicitly given by $ \overline{N}  m'' N',\  \overline{N'}  m' \nu,\  \overline{N'} M_{L'} N',\ \overline{N} M_R C \overline{N}^T$
where $m'$ and $M_{L'}$ also appear at the charged-lepton mass matrix.
$m''$ and $M_{R}$ is found as follows:
\begin{align}
m'' 
&=\frac{v_u}{\sqrt2}
\left(\begin{array}{ccc} 
y'_1 & y'_3 & y'_2  \\
y'_3 & y'_2 & y'_1  \\
y'_2 & y'_1 & y'_3  
\end{array} \right)
\left(\begin{array}{ccc} 
a_n & 0 & 0  \\
0 & b_n & 0  \\
0 &0 & c_n  
\end{array} \right), \\
%%%
M_R
&=
m_1 Y^{(1)}_1
\left(\begin{array}{ccc} 
1 &0 &0  \\
0 & 0 & 1  \\
0 & 1 &0 
\end{array} \right)
+
m_2 Y^{(1)}_{1'}
\left(\begin{array}{ccc} 
0 & 0 & 1  \\
0 & 1 &0 \\
1 &0 &0  
\end{array} \right)
+
m_3
\left(\begin{array}{ccc} 
2 y'_1 & -y'_3 & -y'_2  \\
-y'_3 & 2 y'_2 & -y'_1 \\
-y'_2 & -y'_1 & 2y'_3  
\end{array} \right).
\label{eq:neut-masses}
\end{align}
For our convenience, we redefine $m''\equiv \frac{a_n v_u}{\sqrt2}\tilde{m}''$ and $M_R\equiv m_3\tilde{M}_R$.
The resultant neutral fermion mass matrix in basis of $[\nu, N',{N'^c}, {N^c}]^T$ is found as follows:
\begin{align}
 \left(\begin{array}{cccc} 
0 & 0 & m'^T & 0 \\
0&0 & M_{L'}^T & m''^T \\
m' &  M_{L'} & 0& 0 \\
0 & m'' & 0 & M_R
\end{array} \right).
\label{eq:neut-lep1}
\end{align}
When we impose the following mass hierarchies $m' ,m'' \ll  M_{L'}\lesssim M_R$, we find the following neutrino mass matrix:
\begin{align}
m_\nu &\simeq -(M^{-1}_{L'} m')^T m''^T M_R^{-1} m'' (M^{-1}_{L'} m')\nn \\
&=-\frac{(a_n f_1{v_\varphi} v_u)^2}{4 M_1^2 m_3}(\tilde M^{-1}_{L'} \tilde m')^T \tilde m''^T \tilde M_R^{-1} \tilde m'' (\tilde M^{-1}_{L'} \tilde m')\\
&\equiv  \kappa \tilde m_\nu,
\label{eq:neut-lep2}
\end{align}
where $\kappa\equiv -\left(\frac{f_1{v_\varphi}}{ M_1}\right)^2 \frac{(a_n v_u)^2}{4 m_3}$.
% $\kappa\equiv -\frac{(a_n f_1v' v_u)^2}{4 M_1^2 m_3}$.
Remarkably, we obtain new type of seesaw formula where the active neutrino mass matrix is suppressed by inverse of three powers of heavy neutral fermion mass parameters $M_1^{-2} m_3^{-1}$. 
By this formula, we expect the scale of heavy fermions as $\mathcal{O}(100)$ TeV $\lesssim \{M_{1}, m_3\} \lesssim \mathcal{O}(1000)$ TeV when VEVs, $\{v_u, {v_\varphi}\}$, are electroweak scale and dimensionless parameters are $\mathcal{O}(1)$. As some dimensionless parameters would be smaller than $\mathcal{O}(1)$ our heavy fermions would be lighter, and direct test of the model at collider experiments can be expected via heavy fermion production.  
{For example, new charged leptons can be produced by hadron collider as $pp \to \overline{E'} E' $ where $E'$ decays via Yukawa coupling such as $E' \to H^- \nu/ H(h) \ell^-$; $H^-$ and $H$ are charged Higgs and heavy neutral Higgs. The signals from $E'$ productions could be searched for at the LHC experiments~\cite{ATLAS:2023sbu, CMS:2024bni}.}

%\noindent{\bf \underline{Non-unitarity}}: \\
Before further discussing the neutrino sector, 
we briefly mention a non-unitarity matrix $U'_{\nu}$ since we have larger mass matrix than the canonical seesaw and the mixing matrix to diagonalize our active neutrino mass matrix would be deviate from the unitarity.
This is typically parametrized by the form 
\begin{align}
U'_{\nu}\equiv \left(1-\frac12 F^\dag F\right) U_{\nu},
\end{align}
where $U_\nu$ is defined by a unitary matrix and $\frac12 F^\dag F$ represents deviation from the unitarity.
Then, our $F$, which is a hermitian matrix, is found as $M_{L'}^{-1} m'$.
Applying global constraints~\cite{Fernandez-Martinez:2016lgt} for the above formula,
the following constraints are obtained~\cite{Agostinho:2017wfs}
\begin{align}
|F^\dag F|\le  
\left[\begin{array}{ccc} 
2.5\times 10^{-3} & 2.4\times 10^{-5}  & 2.7\times 10^{-3}  \\
2.4\times 10^{-5}  & 4.0\times 10^{-4}  & 1.2\times 10^{-3}  \\
2.7\times 10^{-3}  & 1.2\times 10^{-3}  & 5.6\times 10^{-3} \\
 \end{array}\right].
\end{align} 
%%%
$|F^\dag F|$ can be rewritten by $\frac{|f_1 {v_\varphi}|^2}{|M_1|^2} |\tilde M_{L'}^{-1} \tilde m'|^2$.
When the scale of $|\tilde M_{L'}|$ is the same as the one of $|\tilde m'|$, we need $\frac{|f_1 {v_\varphi}|} {|M_1|} \lesssim {\cal O}(10^{-3})$ at most in order to satisfy the above conditions. 
%In our numerical analysis below, we implicitly require these bounds.
This bound partially affects $\kappa$ which is found as
\begin{align}
|\kappa| \le    {\cal O}(10^{-6}) \times \left|\frac{(a_n v_u)^2}{4 m_3}\right|. \label{eq:unitarity}
\end{align}
%We find there still exist several degrees of freedom.
Furthermore, $\kappa$ is written in terms of the observed atmospheric mass squared difference $\Delta m^2_{\rm atm}$ and scaleless neutrino mass eigenvalues $\tilde D_\nu$. 
In order to formulate it, we define the mass values as follows. 
$\tilde m_\nu$ is diagonalized by $\tilde D_\nu \simeq U_\nu^T \tilde m_\nu U_\nu$~\footnote{Here, we assume $U'_\nu\sim U_\nu$ so that it satisfies the unitarity $U_\nu^\dag U_\nu=1$.}, where $\tilde D_\nu={\rm diag}[\tilde D_{\nu_1},\tilde D_{\nu_2},\tilde D_{\nu_3}]$,
where NH(IH) stands for Normal(Inverted) Hierarchy.
Then, we find 
\begin{align} 
({\rm NH)}:\ \kappa &= \frac{\Delta m^2_{\rm atm}}{\tilde D_{\nu_3}^2 - \tilde D_{\nu_1}^2}, \quad
({\rm IH)}:\ \kappa = \frac{\Delta m^2_{\rm atm}}{\tilde D_{\nu_2}^2 - \tilde D_{\nu_3}^2}.\label{eq:kappa}
\end{align}
Solar mass square difference $\Delta m^2_{\rm sol}$ is found to be
\begin{align} 
 \Delta m^2_{\rm sol} &=\kappa(\tilde D_{\nu_2}^2 - \tilde D_{\nu_1}^2)  , 
\end{align}
where $\kappa$ depends on the hierarchy in Eq.~(\ref{eq:kappa}).
Sum of the neutrino mass eigenvalues $\sum D_\nu\equiv \kappa(\tilde D_{\nu_1}+\tilde D_{\nu_2}+\tilde D_{\nu_3})$ is constrained by some experiments;
$\sum D_\nu\le120$ meV from the minimal standard cosmological model~\cite{Vagnozzi:2017ovm, Planck:2018vyg}~\footnote{The upper bound becomes weaker if the data are analyzed in the context of extended cosmological models~\cite{ParticleDataGroup:2014cgo}.} 
and $\sum D_\nu\le72$ meV from the combined result from DESI and CMB~\cite{DESI:2024mwx}.
We adopt the standard parametrization of the lepton mixing matrix $U \equiv V_{e_L}^\dag U_\nu$ where the Majorana phase is defined by $[1,e^{i\alpha_{21}/2},e^{i\alpha_{31}/2}]$~\cite{Okada:2019uoy}. Then, the Majorana phases are found as 
\begin{align}
\text{Re}[U^*_{e1} U_{e2}] = c_{12} s_{12} c_{13}^2 \cos \left( \frac{\alpha_{21}}{2} \right), \
 \text{Re}[U^*_{e1} U_{e3}] = c_{12} s_{13} c_{13} \cos \left( \frac{\alpha_{31}}{2} - \delta_{CP} \right), \\
 %%%
 \text{Im}[U^*_{e1} U_{e2}] = c_{12} s_{12} c_{13}^2 \sin \left( \frac{\alpha_{21}}{2} \right), \
 \text{Im}[U^*_{e1} U_{e3}] = c_{12} s_{13} c_{13} \sin \left( \frac{\alpha_{31}}{2} - \delta_{CP} \right),
\end{align}
where $\alpha_{21}/2,\ \frac{\alpha_{31}}{2} - \delta_{CP}$
are subtracted from $\pi$, when $\cos(\alpha_{21}/2)$ and $\cos \left( \frac{\alpha_{31}}{2} - \delta_{CP} \right)$ are negative.
The effective neutrino mass for neutrinoless double beta decay $\langle m_{ee} \rangle$ is also restricted by the current KamLAND-Zen data;
$\langle m_{ee}\rangle<(28-122)$ meV at 90 \% confidence level~\cite{KamLAND-Zen:2024eml}.
%$\langle m_{ee}\rangle<(36–156)$ meV at 90 \% confidence level~\cite{KamLAND-Zen:2022tow}  
where it is defined by
\begin{align}
\langle m_{ee}\rangle=\kappa|\tilde D_{\nu_1} \cos^2\theta_{12} \cos^2\theta_{13}+\tilde D_{\nu_2} \sin^2\theta_{12} \cos^2\theta_{13}e^{i\alpha_{2}}+\tilde D_{\nu_3} \sin^2\theta_{13}e^{i(\alpha_{3}-2\delta_{CP})}|.
\end{align}

\section{Numerical analysis and phenomenology}
\label{sec:III}
In this section,  we perform the chi square numerical analysis.
At first we randomly select the values of our input parameters within the following ranges:
\begin{align}
[\tilde f_2,\ \tilde f_3,\ \tilde M_2,\ \tilde M_3,\ \tilde b_n,\ \tilde c_n,\ \tilde m_1,\ \tilde m_2]\in [10^{-3}-10^3],
\end{align}
where $\tilde f_{2,3}\equiv f_{2,3}/f_1$, $\tilde M_{2,3}\equiv M_{2,3}/M_1$, $\tilde b_n(\tilde c_n)\equiv b_n(c_n)/a_n$,
and $\tilde m_{1,2}\equiv m_{1,2}/m_3$. Without loss of generality, the above parameters can be real by phase redefinitions.
{Note that, in our scenario of $m' ,m'' \ll  M_{L'}\lesssim M_R$, new lepton masses are mostly determined by the scale of $M_{L'}$ and $M_R$.
The masses of $L'$ including neutral and charged component are given by $M_{L'}$ while that of $N$ are determined by $M_R$.  
Thus parameters $[ \tilde M_2,\ \tilde M_3, \ \tilde m_1,\ \tilde m_2]$ are related to mass hierarchy among generations of $L'$ and $N$. 
Taking $\{M_1, m_3 \} \sim {\rm few} \times 100$ TeV scale as discussed below {Eq.~(11)}, new lepton masses can be ${\rm few} \times 100$ GeV to ${\rm} {\rm few} \times 100$ PeV for $[ \tilde M_2,\ \tilde M_3, \ \tilde m_1,\ \tilde m_2] \in [10^{-3}-10^3]$. The other dimensionless parameters are relevant for active neutrino masses. }

The modulus $\tau$ is randomly selected in the fundamental region.
%%%
We adopt five reliable observables $\Delta m^2_{\rm atm},\ \Delta m^2_{\rm sol},\ s_{12},\ s_{23},\ s_{13}$ as our chi square analysis
where we refer to Nufit6.0~\cite{Esteban:2024eli}.
{
 We then estimate the $\Delta \chi^2$ value using the formula of 
\begin{equation}
 \Delta \chi^2 = \sum_{i}  \left( \frac{O_i^{\rm obs} - O_i^{\rm th}}{\delta O_i^{\rm exp}} \right)^2, \label{eq:chi-square}
\end{equation}
where $O_i^{\rm obs (th)}$ is the observed (theoretically obtained) value of observables, $\delta O_i^{\rm exp}$ is corresponding experimental error and 
$i$ distinguishes different observables.
}
The number of  degrees of freedom (dof) is fourteen,
 therefore twelve real parameters of 
$a_e, b_e, c_e$ (only for charged-leptons),  $ \tilde f_{2,3},\ \tilde M_{2,3}$, $\tilde b_n,\ \tilde c_n,\ \tilde m_1,\ \tilde m_2,\ \kappa$    
(only for the neutrino sector), and complex one of  $\tau$. 
%Thus, usual models would not have any clear predictions.
{However,
five parameters $\tau,\ \tilde f_{2,3},\ \tilde M_{2,3}$ contribute to both of the charged-lepton and active neutrino sectors.
Through our analysis, we adopt the allowed regions only when the chi square should be within 5 $\sigma$ confidence level (that is equivalent to $\Delta \chi^2\approx 37.1$(dof$=$5) {with dof=5 being number of observables in neutrino sector} ). 
In addition, we restrict ourselves to focus on two fixed points of $\tau$; $i$ and $\omega(\equiv e^{2\pi i/3})$ which are favored by a type IIB string compactifications with background fluxes~\cite{Ishiguro:2020tmo}.
{The fixed points and the nearby region are statistically favored from moduli stabilization in flux compactification scheme. }}
%In the top-down approach of Ref. [30], distributions of moduli fields from a finite number of string flux vacua suggest us that the moduli VEVs are statistically favored at the other special points

\subsection{NH}
At first, we discuss the case of NH.
{We show some predicted values of neutrino observables for the parameter points satisfying neutrino data where we also find the points satisfy Eq.~\eqref{eq:unitarity} for unitarity constraint. }

\subsubsection{$\tau=\omega$}
%
 %%%%%%%%%%%%%%%%%%%
\begin{figure}[tb]
\begin{center}
\includegraphics[width=50.0mm]{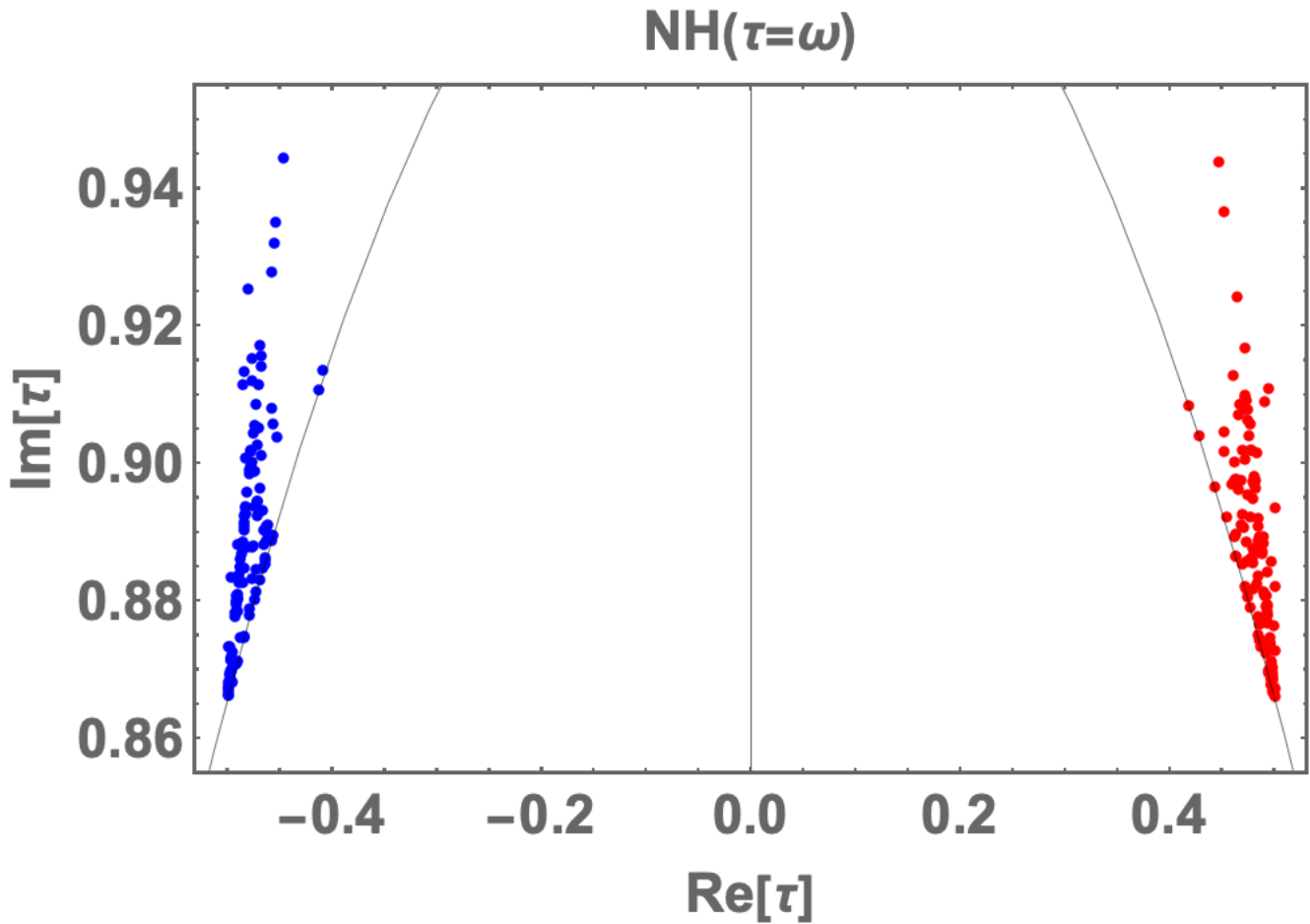} \quad
%%%
\includegraphics[width=50.0mm]{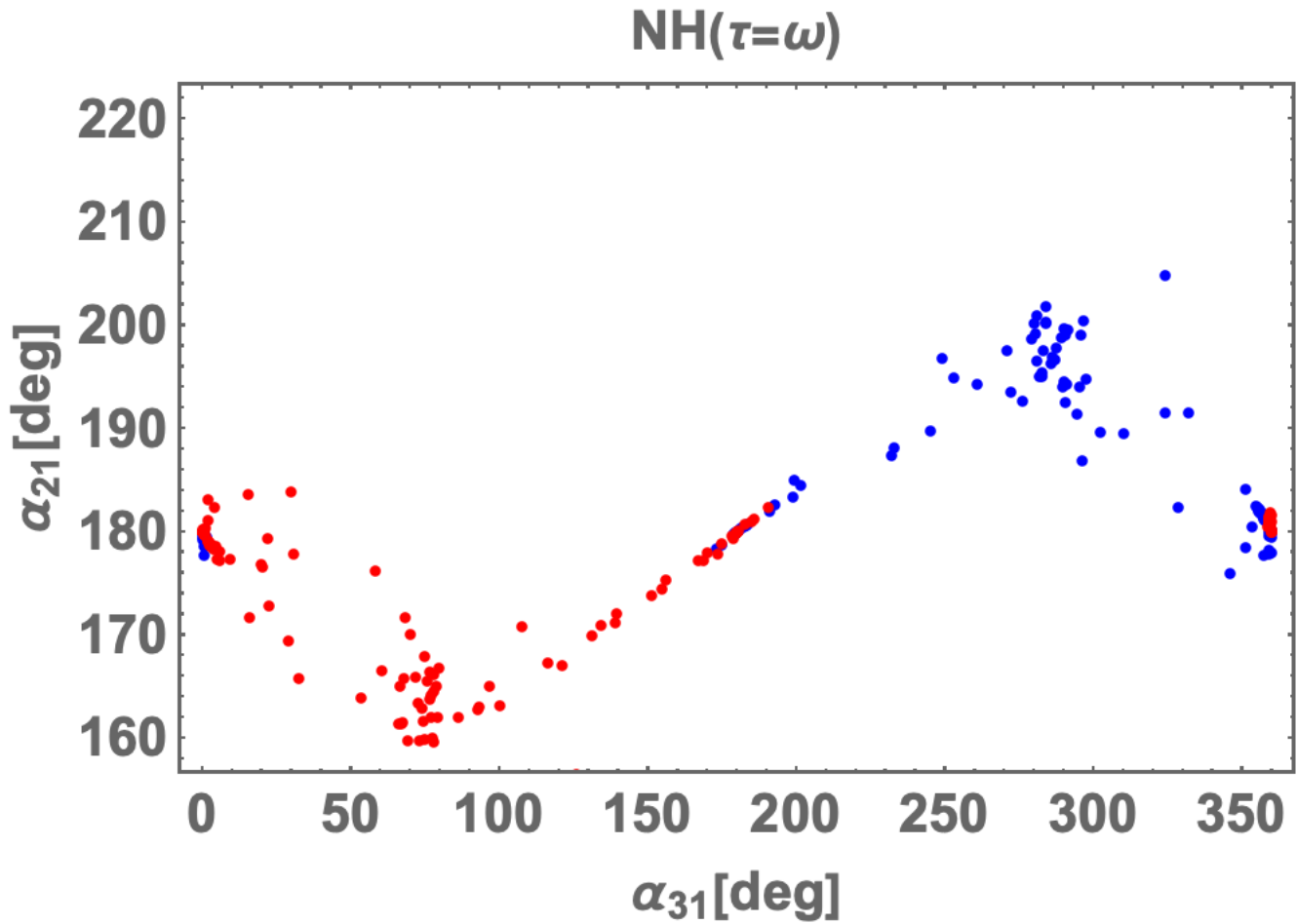} \quad
%%%
\caption{Allowed regions of $\tau$ in the fundamental region (left) and Majorana phases (right), where blue plots represent the one at nearby $\tau=\omega$ and red one $\tau=-\omega^*$. }
  \label{fig:omega_nh1}
\end{center}\end{figure}
%%%%%%%%%%%%%%%%%%%   
Fig. \ref{fig:omega_nh1} shows the allowed region of $\tau$ in the fundamental region (left) and Majorana phases (right), where blue plots represent the one at nearby $\tau=\omega$ and red one $\tau=-\omega^*$.
We find Majorana phases have following tendency $175^\circ\lesssim \alpha_{21}\lesssim 205^\circ$ and $200^\circ\lesssim \alpha_{31}\lesssim 360^\circ$.
% The remaining space is occupied by the allowed region at nearby $\tau=-\omega^*$.

 %%%%%%%%%%%%%%%%%%%
\begin{figure}[tb]
\begin{center}
\includegraphics[width=70.0mm]{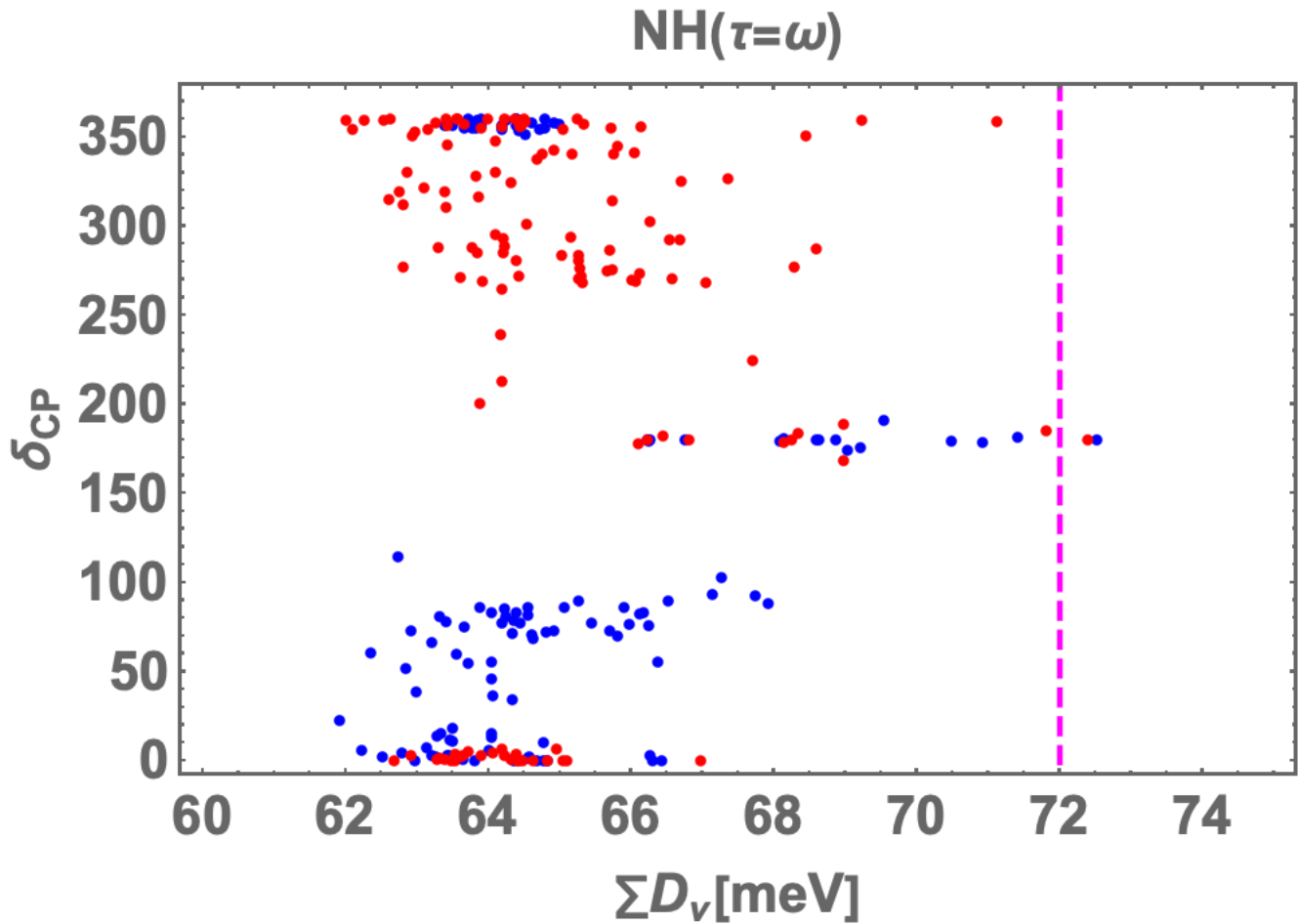} \quad
%%%
\includegraphics[width=70.0mm]{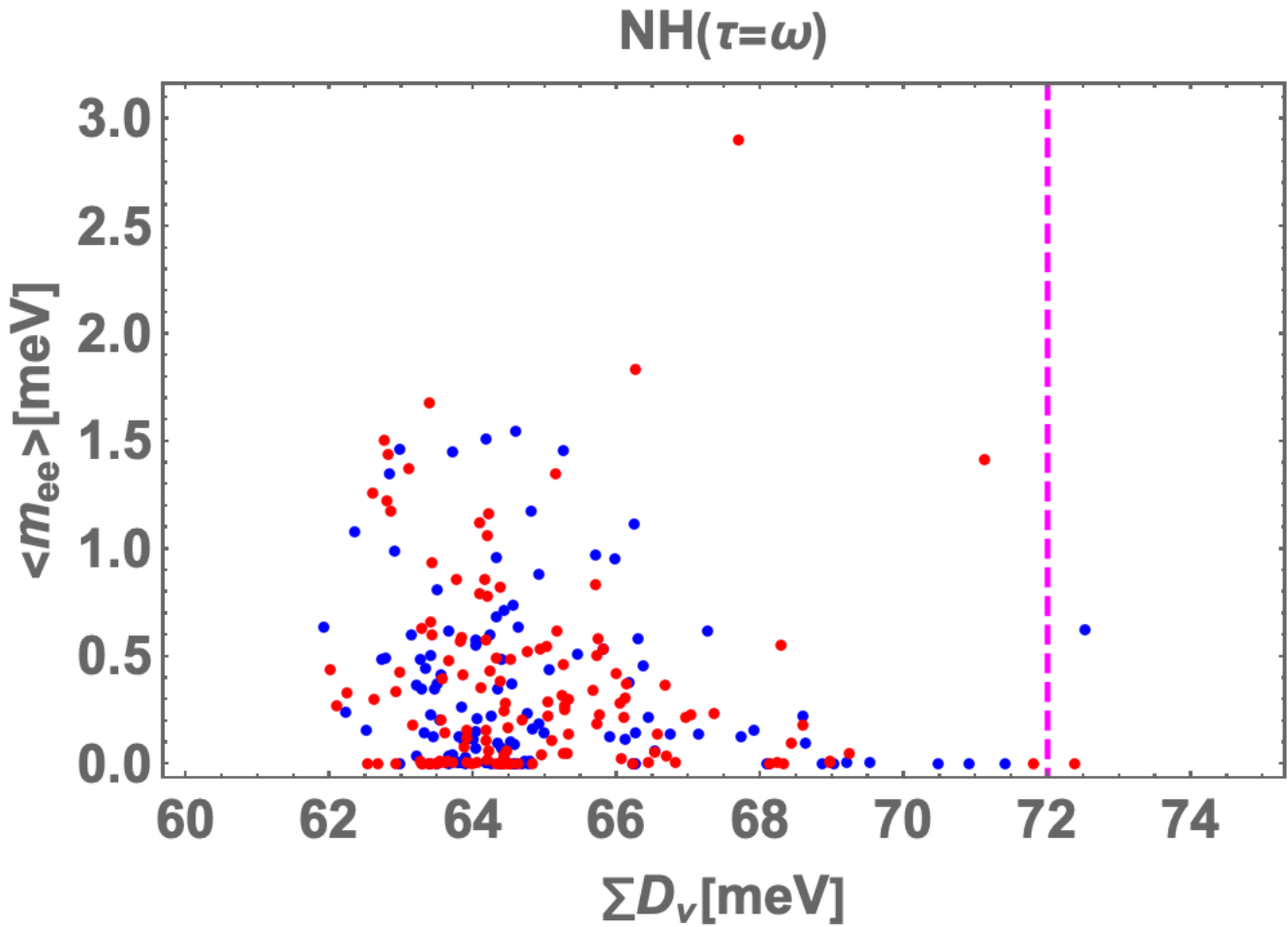} \quad
%%%
%\includegraphics[width=50.0mm]{figs/m1-mee_nh-omega.pdf} 
\caption{Allowed regions of $\delta_{CP}$(left) and  $\langle m_{ee}\rangle$(right) in terms of $\sum D_\nu$ meV,  
where the color legends of plots are the same as the one in Fig.~\ref{fig:omega_nh1}.
The magenta vertical dotted line at $\sum D_\nu=72$ meV represents the bound on DESI and CMB combined result. }
  \label{fig:omega_nh2}
\end{center}\end{figure}
%%%%%%%%%%%%%%%%%%%   
%
Fig.~\ref{fig:omega_nh2} shows the allowed regions of $\delta_{CP}$(left) and  $\langle m_{ee}\rangle$(right) in terms of $\sum D_\nu$ meV,  
%and$\langle m_{ee}\rangle$ in terms of  (right), 
where the color legends of plots are the same as the one in Fig.~\ref{fig:omega_nh1}.
The magenta vertical dotted line at $\sum D_\nu=72$ meV represents the bound on DESI and CMB combined result.
These figures tell us that the tendency of $\delta_{CP}$ is [$0^\circ-180^\circ$] and  $\sum D_\nu$ is [$62-72.5$] meV.
The maximum value of  $\sum D_\nu$  is located at nearby the bound of DESI and CMB combined result and its testability would be good.
On the other hand, the allowed region of $\langle m_{ee}\rangle$ is [$0-3$] meV that is much lower than the current lower bound of KamLAND-Zen data.

\subsubsection{$\tau=i$}
%
%
 %%%%%%%%%%%%%%%%%%%
\begin{figure}[tb]
\begin{center}
\includegraphics[width=50.0mm]{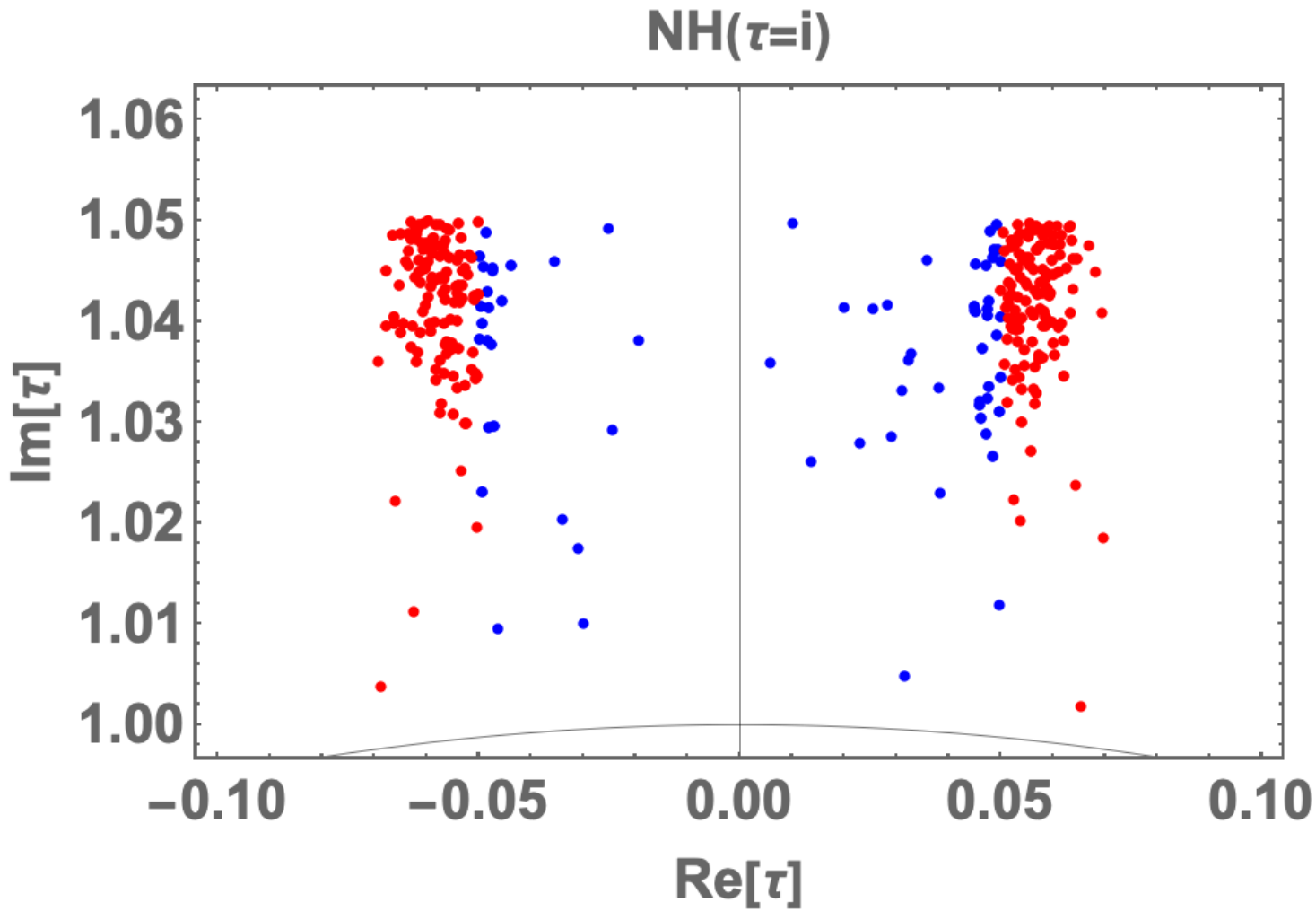} \quad
%%%
\includegraphics[width=50.0mm]{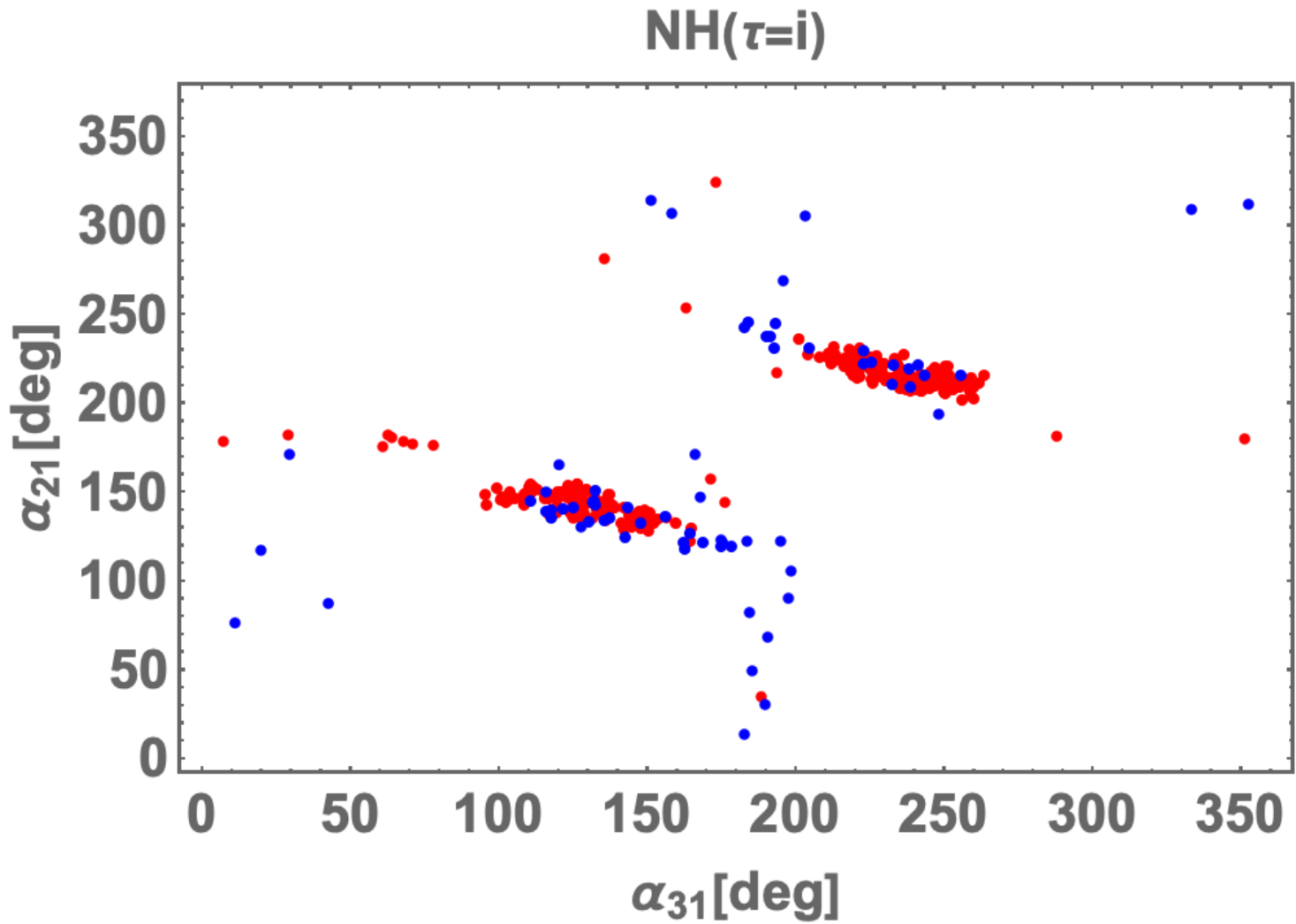} \quad
%%%
\caption{Allowed regions of $\tau$ in the fundamental region (left) and Majorana phases (right), where blue plots represent $|Re[\tau]|\le0.05$ and red ones $0.05< |Re[\tau]|\le0.07$. }
  \label{fig:i_nh1}
\end{center}\end{figure}
%%%%%%%%%%%%%%%%%%%   
%
Fig. \ref{fig:i_nh1} shows the allowed regions of $\tau$ in the fundamental region (left) and Majorana phases (right), where blue plots represent $|Re[\tau]|\le0.05$ and red ones $0.05< |Re[\tau]|\le0.07$.
Majorana phases run over whole the ranges but there would exist a correlation between these two phases.

 %%%%%%%%%%%%%%%%%%%
\begin{figure}[tb]
\begin{center}
\includegraphics[width=70.0mm]{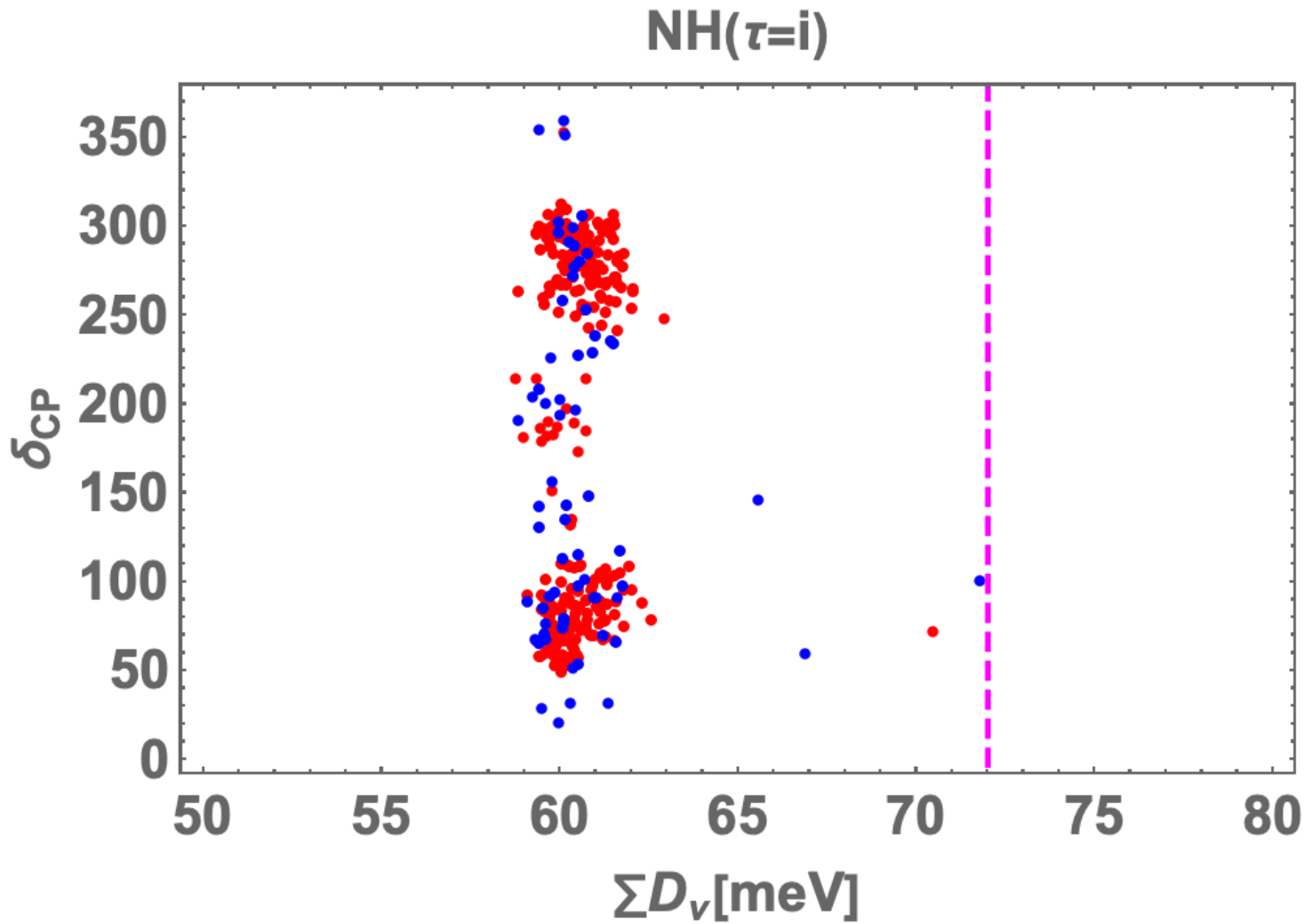} \quad
%%%
\includegraphics[width=70.0mm]{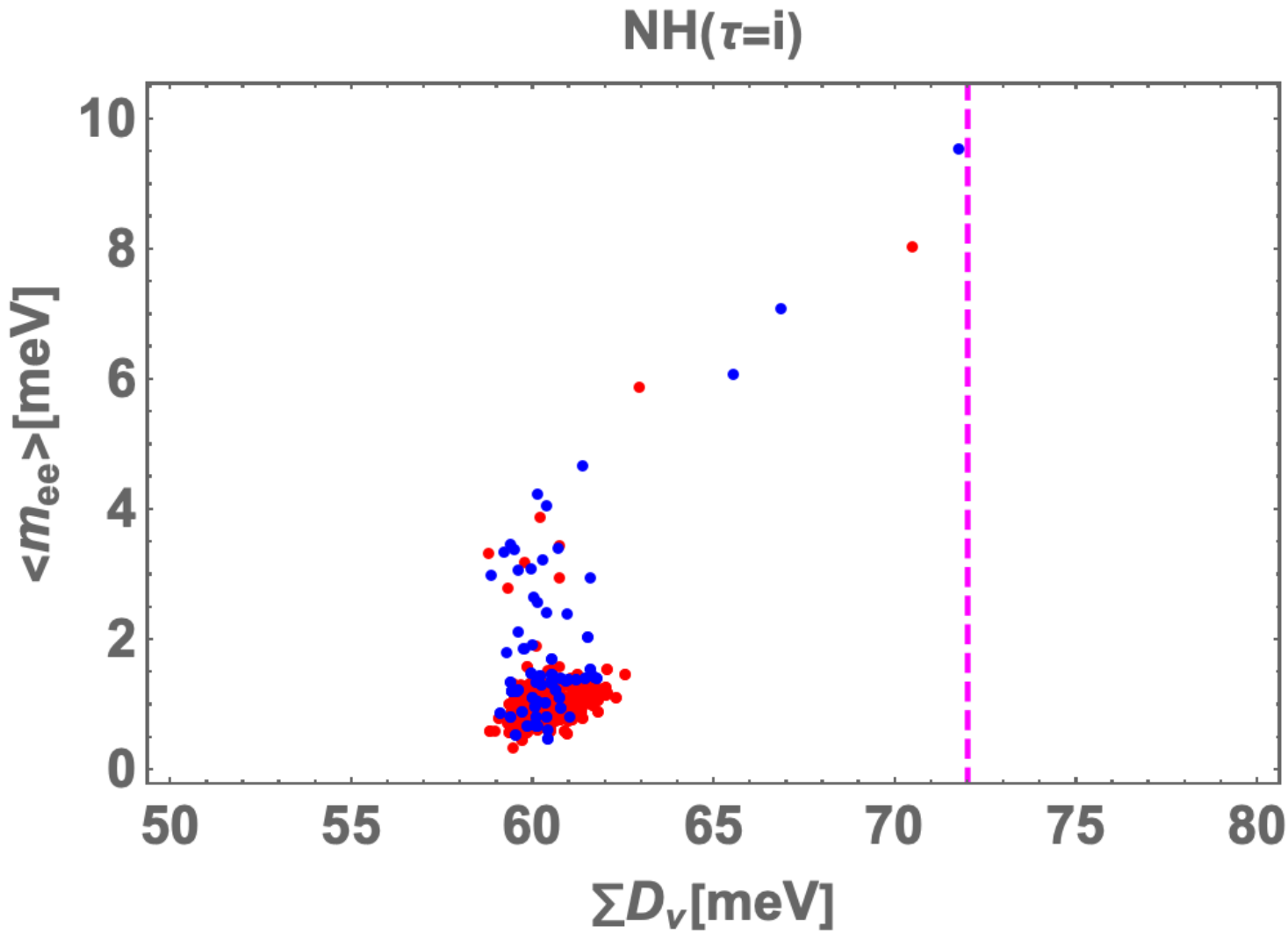} \quad
%%%
%\includegraphics[width=50.0mm]{figs/m1-mee_nh-i.pdf} 
\caption{Allowed regions of $\delta_{CP}$(left) and  $\langle m_{ee}\rangle$(right) in terms of $\sum D_\nu$ meV,  
where the color legends of plots and the magenta vertical dotted line are the same as the one in Fig.~\ref{fig:omega_nh1}. }
  \label{fig:i_nh2}
\end{center}\end{figure}
%%%%%%%%%%%%%%%%%%%   
%
%
Fig.~\ref{fig:i_nh2} shows the allowed regions of $\delta_{CP}$(left) and  $\langle m_{ee}\rangle$(right) in terms of $\sum D_\nu$ meV,  
%and$\langle m_{ee}\rangle$ in terms of  (right), 
where the color legends of plots and the magenta vertical dotted line are the same as the one in Fig.~\ref{fig:omega_nh1}.
%The magenta vertical dotted line at $\sum D_\nu=72$ meV represents the bound on DESI and CMB combined result.
%
$\delta_{CP}$ runs whole the region, but  $\sum D_\nu$ tends to be localized at nearby $60$ meV.
The maximum value of  $\sum D_\nu$  is located at nearby the bound of DESI and CMB combined result and its testability would be good.
On the other hand, the allowed region of $\langle m_{ee}\rangle$ is [$0-10$] meV that is lower than the current lower bound of KamLAND-Zen data.

\subsection{IH}
Second, we discuss the case of IH.
{We show some predicted values of neutrino observables for the parameter points satisfying neutrino data where we also find the points satisfy Eq.~\eqref{eq:unitarity} for unitarity constraint. }

\subsubsection{$\tau=\omega$}

 %%%%%%%%%%%%%%%%%%%
\begin{figure}[tb]
\begin{center}
\includegraphics[width=70.0mm]{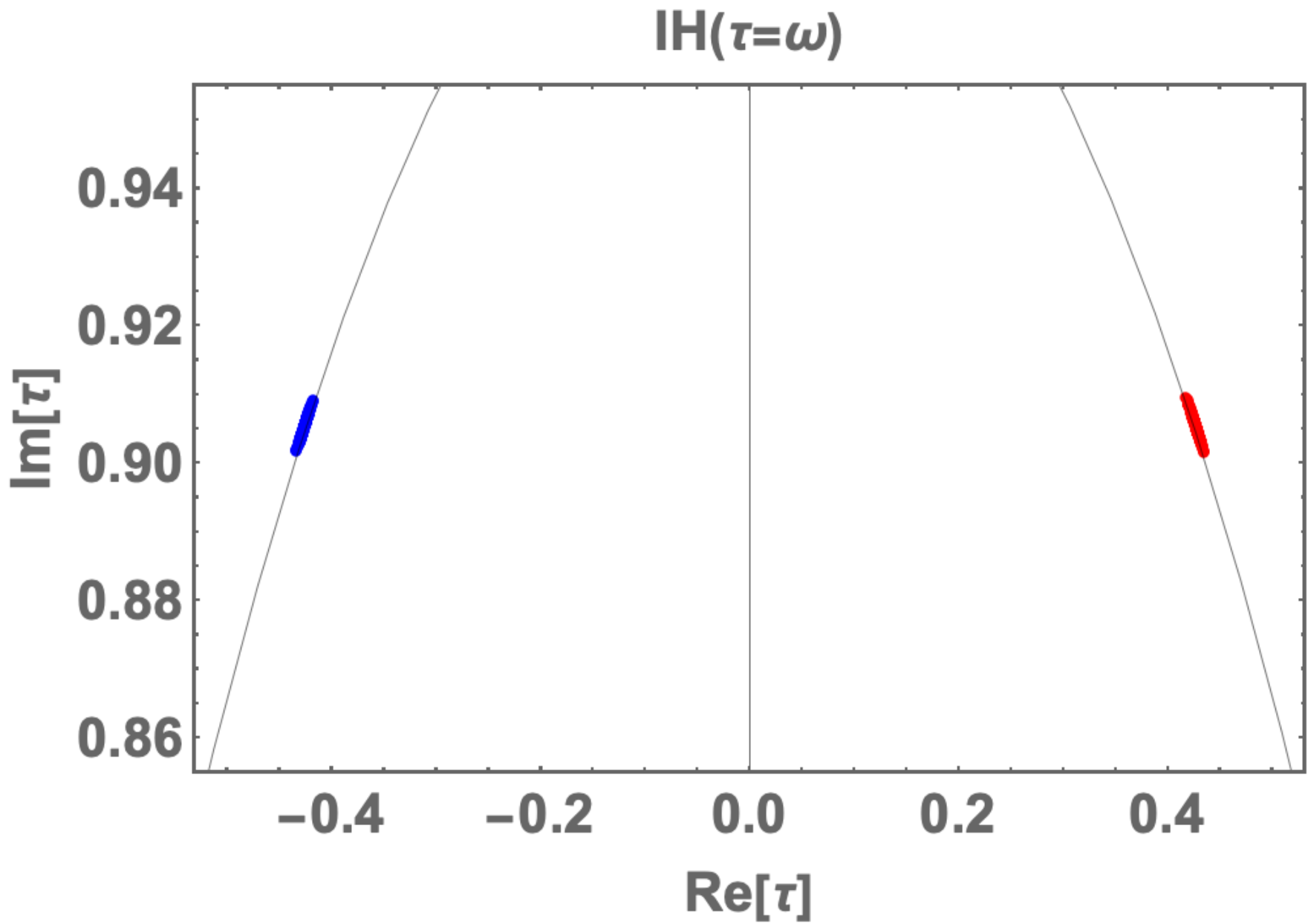} \quad
%%%
\includegraphics[width=70.0mm]{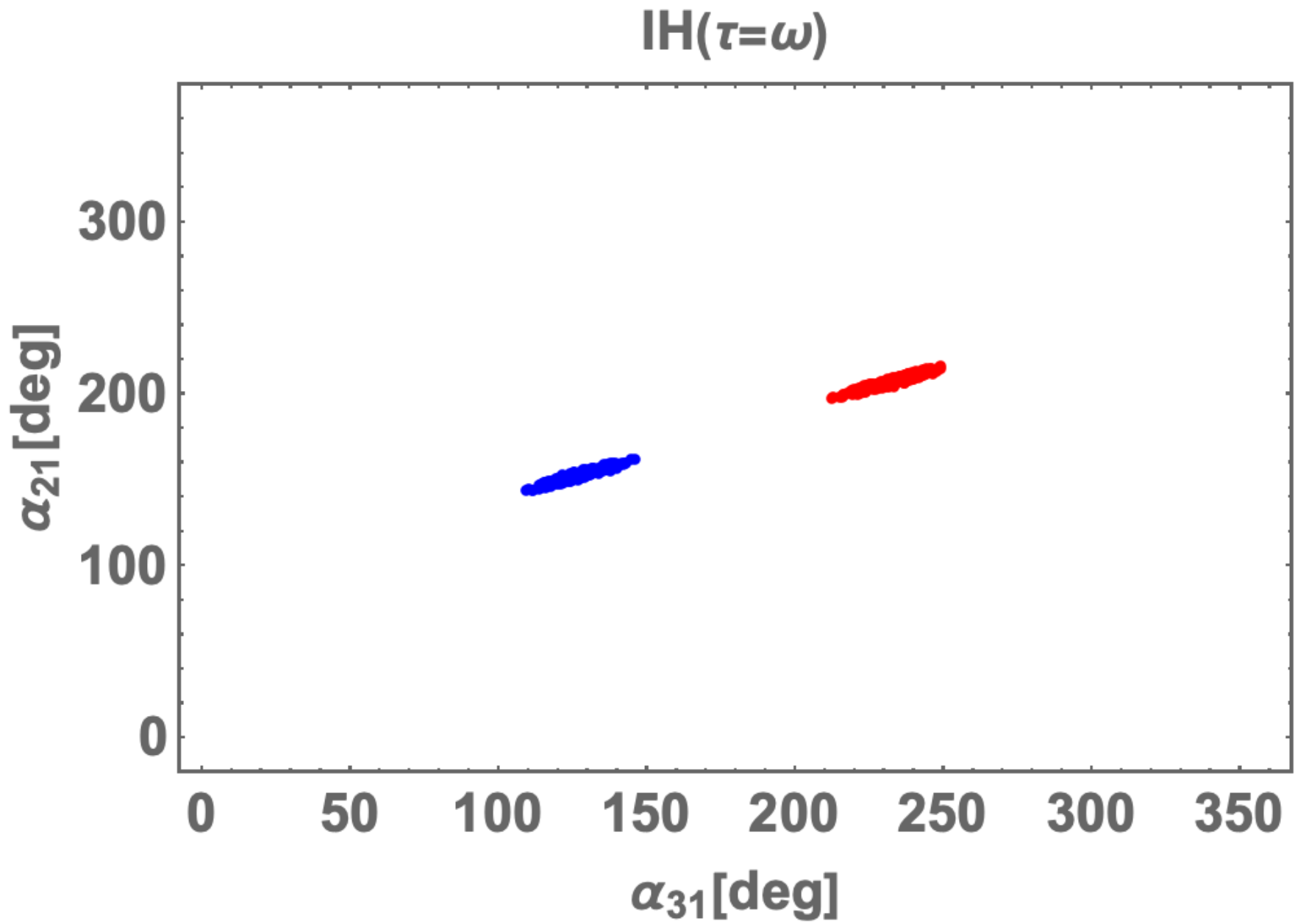} \quad
%%%
\caption{Allowed regions of $\tau$ in the fundamental region (left) and Majorana phases (right), where the color legends are the same as the one of Fig.~\ref{fig:omega_nh1}.}
  \label{fig:omega_ih1}
\end{center}\end{figure}
%%%%%%%%%%%%%%%%%%%   
%
Fig.~\ref{fig:omega_ih1} show the allowed regions of $\tau$ in the fundamental region (left) and Majorana phases (right), where the color legends are the same as the one of Fig.~\ref{fig:omega_nh1}.
Majorana phases have a unique tendency that both of phases are localized at $140^\circ\lesssim \alpha_{21}\lesssim 170^\circ$ and $100^\circ\lesssim \alpha_{31}\lesssim 150^\circ$.

 %%%%%%%%%%%%%%%%%%%
\begin{figure}[tb]
\begin{center}
\includegraphics[width=70.0mm]{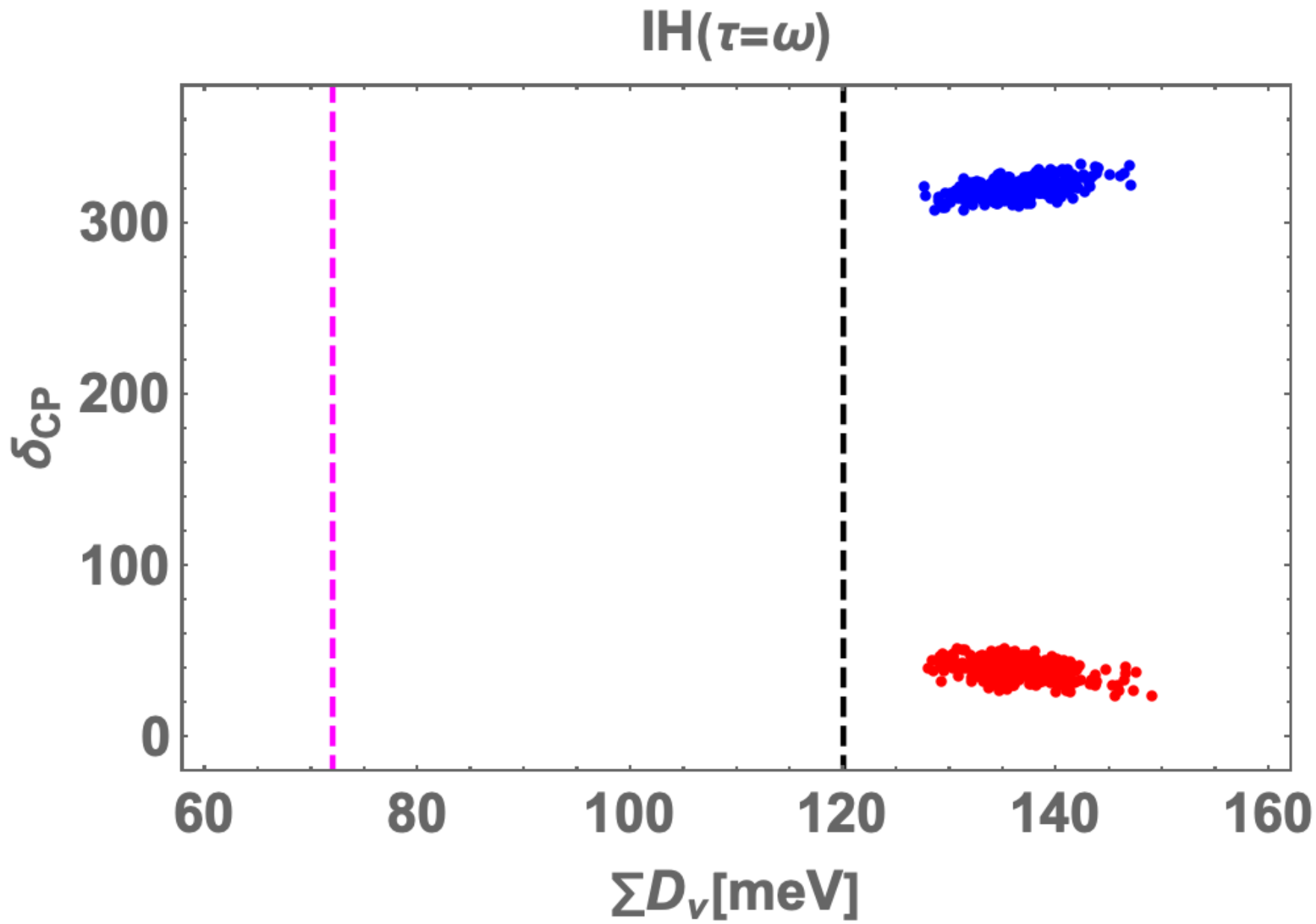} \quad
%%%
\includegraphics[width=70.0mm]{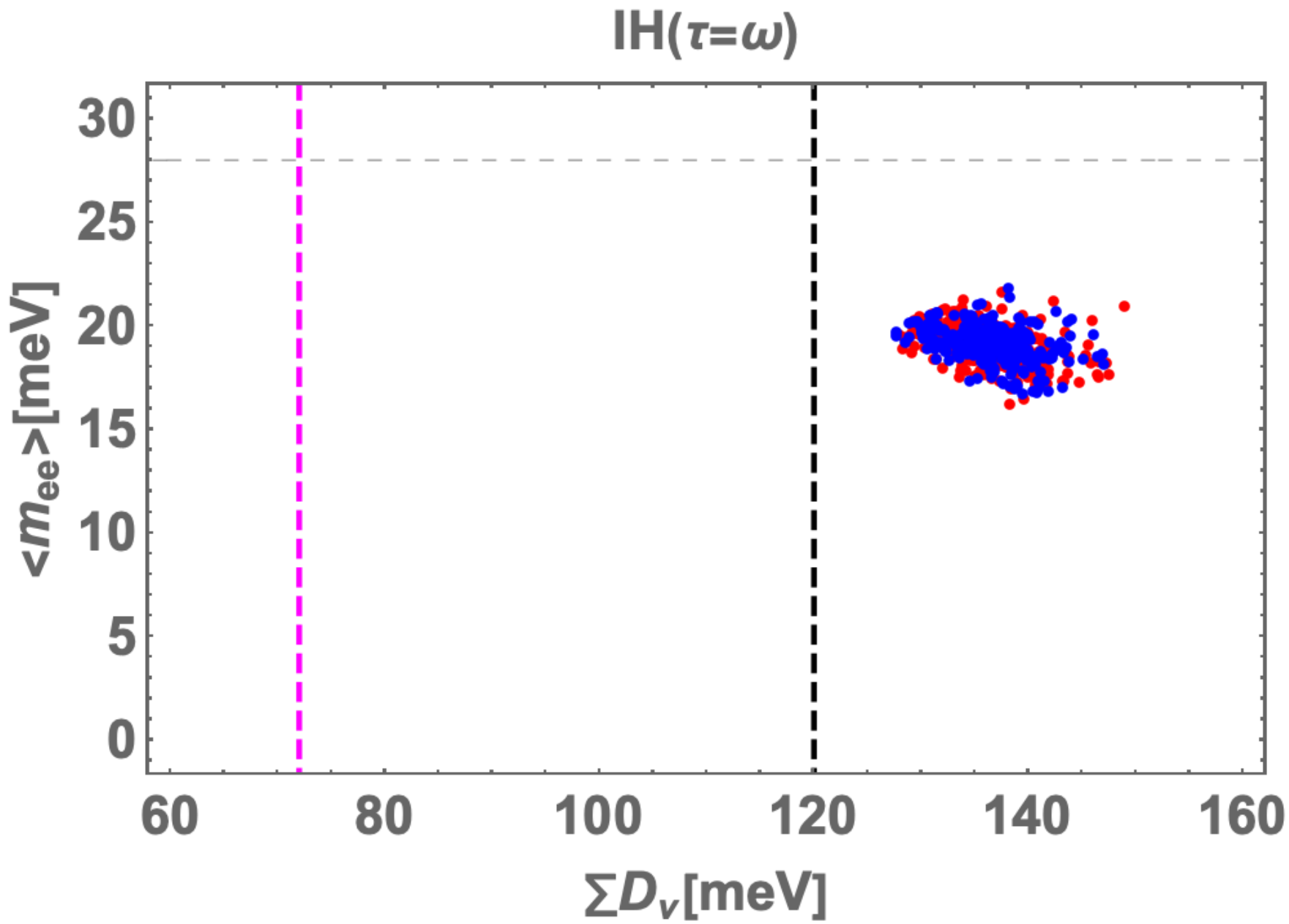} %\quad
%%%
%\includegraphics[width=50.0mm]{figs/m3-mee_ih-omega.pdf} 
\caption{ Allowed regions of $\delta_{CP}$(left) and  $\langle m_{ee}\rangle$(right) in terms of $\sum D_\nu$ meV,  
where the color legends of plots and the magenta vertical dotted line are the same as the one in Fig.~\ref{fig:omega_nh1}. 
The black vertical dotted line at $\sum D_\nu=120$ meV represents the bound  from the minimal
standard cosmological model. The horizontal gray dotted line ($28$ meV) is lower bound of KamLAND-Zen data.}
  \label{fig:omega_ih2}
\end{center}\end{figure}
%%%%%%%%%%%%%%%%%%%   
%
Fig.~\ref{fig:omega_ih2} shows the allowed regions of $\delta_{CP}$(left) and  $\langle m_{ee}\rangle$(right) in terms of $\sum D_\nu$ meV,  
%and$\langle m_{ee}\rangle$ in terms of  (right), 
where the color legends of plots and the magenta vertical dotted line are the same as the one in Fig.~\ref{fig:omega_nh2}.
The black vertical dotted line at $\sum D_\nu=120$ meV represents the bound  from the minimal
standard cosmological model and the horizontal gray dotted line ($28$ meV) is lower bound of KamLAND-Zen data.
$\delta_{CP}$ is localized at the region of $[300^\circ-330^\circ]$, and $\sum D_\nu$ is localized at the range of  $[125-150]$ meV.
Even though  $\sum D_\nu$ does not satisfy the cosmological bound, it would easily be relax this bound in case of IH.
%The maximum value of  $\sum D_\nu$  is located at nearby the bound of DESI and CMB combined result and its testability would be good.
The allowed region of $\langle m_{ee}\rangle$ is [$15-22$] meV that would reach at the lower bound of KamLAND-Zen data soon.

\subsubsection{$\tau=i$}
We have no solutions below $\Delta \chi^2=10^6$ in this case.
Thus, we conclude that our model disfavors the case of IH with $\tau=i$.

%%%%%%%%%%%%%%%%%%%%
\section{Summary and discussion}
\label{sec:IV}
We have proposed a new type of lepton seesaw model introducing a modular $A_4$ flavor symmetry in which isospin doublet vector fermions play an important role in constructing seesaw mechanisms for both the charged-lepton mass matrix and the neutrino one.
%
%The charged-lepton mass matrix is induced via two by two block
%
In particular, we have realized a new type of seesaw formula in the generation of active neutrino masses in which cubic of heavy fermion masses appears in the denominator. From the formula, we find that the heavy fermion mass scale is at most $\mathcal{O}(100)$ TeV to $\mathcal{O}(1000)$ TeV when the dimensionless parameters are $\mathcal{O}(1)$ and scalar VEVs are electroweak scale. As some dimensionless parameters would be smaller than $\mathcal{O}(1)$ our heavy fermions would be lighter, and direct test of the model at collider experiments can be expected via heavy fermion production. 
{For example, heavy charged lepton would be produced at hadron collider as $pp \to \overline{E'}E'$ and its signal could be searched for at the LHC experiments. }
Through the chi square analysis, we have some tendencies about observables focusing on two fixed points $\tau=i,\ \omega$.
{We have found some predicted tendencies in our allowed parameter region. For NH($\tau = \omega$), the sum of neutrino mass is mostly lighter than 70 meV and the $\delta_{CP}$ value spans wide range. For NH($\tau=i$), the sum of neutrino mass tends to be small around 60 meV and the $\delta_{CP}$ value also spans widely. In both NH cases, $\langle m_{ee} \rangle$ values are small to be observed. 
For IH($\tau= \omega$), the sum of neutrino mass is within [130, 150] meV and the $\delta_{CP}$ value tends to be around $[300^\circ-330^\circ]$ and $[20^\circ-60^\circ]$. The $\langle m_{ee} \rangle$ value is in [$15$, $22$] meV. These results would distinguish the model from other seesaw models with modular $A_4$. In particular, the sum of neutrino mass and $\langle m_{ee} \rangle$ can distinguish the model from the other lepton seesaw model in ref.~\cite{Nomura:2024ctl} since the latter model predict the values much larger than the current model. Thus more data regarding neutrino observables would test the models. }

%%%%%%%%%%%%%%%%%%%%%%%%%%%%%%%%%%%
\section*{Acknowledgments}
%\vspace{0.3cm}
The work was supported by the Fundamental Research Funds for the Central Universities (T.~N.).
%%%%%%%%%%%%%%%%%%%%%%%%%%%%%%%%%%%

% Ref Style
% Including title
%\bibliographystyle{utphys}
\bibliography{ctma4.bib}
\end{document}